\documentclass{iopjournal}
\usepackage[T1]{fontenc}
\usepackage{lmodern}
\usepackage{amsmath,amssymb,bm}
\usepackage[nopatch=footnote]{microtype}
\usepackage[numbers,sort&compress]{natbib}
\usepackage{url,needspace}
\usepackage{array,graphicx,booktabs,longtable}
\hypersetup{hidelinks,pdftitle={Canonical-Ensemble Response from Spin-Adapted Time-Dependent Density Functional Theory},pdfauthor={Xiaoyu Zhang}}

\newcommand{\doi}[1]{\href{https://doi.org/#1}{\nolinkurl{#1}}}

\makeatletter
\renewcommand\subsubsection{\@startsection{subsubsection}{3}{\z@}%
  {-2.2ex\@plus -0.7ex\@minus -0.2ex}{0.8ex\@plus 0.2ex}%
  {\normalfont\normalsize\bfseries\upshape}}
\makeatother
\allowdisplaybreaks[1]
\AtBeginEnvironment{thebibliography}{\interlinepenalty=10000\setstretch{1.0}}

\begin{document}
\articletype{Paper}
\title{Transition Density Matrix in Spin-Adapted TDDFT}
\author{Xiaoyu Zhang$^{1,2,*}$\orcid{0009-0009-4178-3519}}
\affil{$^1$College of Chemistry and Molecular Engineering, Peking University, Beijing, China}
\affil{$^2$Department of Chemistry, The University of Hong Kong, Hong Kong, China}
\affil{$^*$Author to whom any correspondence should be addressed.}
\email{zhangxiaoyu@connect.hku.hk}
\begin{abstract}
The transition density matrix between different excited states is important for calculating optical response properties and first-order nonadiabatic coupling matrix elements. Spin-adapted TDDFT (SA-TDDFT) has eliminated spin contamination, achieved minimal spin-completeness, and has been found to have important applications in conical intersections, bond dissociations, diradicals, and triradicals. In this work, we propose a unified formulation for calculating the transition density matrix of excited states from SA-TDDFT. This formulation starts from the second-quantized equation-of-motion (EOM), approximates the expansion of the translation operator, and naturally obtains the relaxed and bilinear parts of the matrix. In the closed-shell and random phase approximation (RPA) limit, this formulation is compared with formulations from quadratic response and pseudo-wavefuntion methods respectively. Computational results demonstrate the accuracy of our method and importance of the relaxed term.

\end{abstract}
\section{Introduction}

Recently, Zhang and Wang proposed  the spin-adapted TDDFT (SA-TDDFT).\cite{ZhangWang2026} Its random-phase approximation (RPA) starting point combines Li and Liu's constructions for $S_f=S_i$ and $S_f=S_i+1$\cite{LiLiu2010} with Zhang and Herbert's construction for $S_f=S_i-1$,\cite{ZhangHerbert2015} where $S_i$ and $S_f$ denote the reference and target spins. Both constructions use the tensor equation-of-motion formalism of Rowe and Ngo-Trong.\cite{Rowe1975} SA-TDDFT has been found to have important applications in conical intersections, bond dissociations, diradicals, and triradicals. However, the transition density matrix of SA-TDDFT hasn't been formulated yet.

The one-particle transition density matrix (TDM) provides matrix elements of one-electron operators by contraction with the corresponding integrals. In TDDFT, electric and magnetic transition dipoles obtained from density matrix response theory determine oscillator and rotatory strengths, while transition dipoles between excited states give access to excited-state absorption~\cite{Furche2001}. At the exact level, electric dipole matrix elements between excited states also enter sums over intermediate states for hyperpolarizabilities and two-photon transition amplitudes~\cite{Olsen1985}. TDDFT formulations of these nonlinear optical properties have been developed for dynamic hyperpolarizabilities~\cite{vanGisbergen1997,Salek2002} and two-photon absorption~\cite{Salek2003}. The transition density also connects electronic structure to nonadiabatic dynamics: between exact nondegenerate electronic states, its contraction with the nuclear derivative of the electron--nuclear potential, divided by the energy gap, gives the derivative coupling~\cite{Chernyak2000,LiLiu2014}. Evaluation of couplings between excited states within TDDFT must account for orbital response and, for basis functions attached to the nuclei, their nuclear derivatives~\cite{LiLiu2014}. Transition densities also enter the propagation of electronic state amplitudes and the evolving density in response-reformulated TDDFT, enabling simulations of electronic dynamics driven by laser fields~\cite{Dar2024}.

The construction of TDMs between excited states has developed through both equation of motion (EOM) methods and response theory. Csanak, Taylor, and Tripathy identified these densities in the spectral representation of a three-particle response function~\cite{Csanak1973}, and Csanak subsequently derived their TDHF equations~\cite{Csanak1974}. The EOM treatment of Yeager, Nascimento, and McKoy gave transition moments bilinear in the RPA excitation and deexcitation amplitudes~\cite{Yeager1975}. Dalgaard's derivation from double residues of the quadratic response function showed that additional response terms are required for correctness through first order in the electron--electron interaction~\cite{Dalgaard1982}. The residue formulation was developed for exact and multiconfigurational self-consistent field states by Olsen and J{\o}rgensen~\cite{Olsen1985}, and subsequently for TDDFT using density matrix response theory~\cite{Furche2001}. Later, Li and Liu formulated nonadiabatic couplings using EOM and response theory~\cite{LiLiu2014}, while derivations based on pseudo-wavefunctions identified transition densities containing static orbital relaxation~\cite{Ou2014,Alguire2015}. The static approximation evaluates the response equation at zero frequency rather than at the energy difference between the two states. In TDHF and adiabatic TDDFT, the latter construction can develop spurious divergences when this difference coincides with another excitation energy~\cite{Parker2016}. The numerical importance of the additional contributions depends on the transition. In TDHF, Sengel{\o}v and Oddershede found the full quadratic response correction generally small for strong transitions in the dipole length representation, but dominant for some weak transitions and near avoided crossings~\cite{Sengelov1988}. In TDDFT calculations on perylene diimide dimers, neglecting orbital relaxation produced excited-state absorption spectra up to five times as intense as those including static relaxation, even away from the spurious resonances~\cite{Parker2018}. A recent study of 82 transitions likewise found substantial differences between unrelaxed and relaxed oscillator strengths, with the largest outliers involving strong deexcitation contributions~\cite{Sirucek2026}. In protonated formaldimine, omitting the orbital response contribution represented by the relaxed density correction caused the line integral of the derivative coupling around the $S_1/S_2$ conical intersection to deviate from the expected value of $\pi$~\cite{Ou2015PW}. These results motivate explicit evaluation of orbital relaxation for individual transitions. For open-shell systems, Wang et al. have derived TDMs and analytic nonadiabatic couplings for excitations that conserve total spin in X-TDDFT~\cite{Wang2026}. Here, we calculate transition densities between distinct states and include orbital relaxation in the SA-TDDFT formulation of Zhang and Wang~\cite{ZhangWang2026}. We differentiate the generalized eigenvalue equation with respect to a spin-independent one-electron perturbation, retaining the response of the reference orbitals and excitation operators. The resulting TDM separates into a bilinear contribution and an orbital relaxation contribution obtained by solving a linear equation involving the orbital Hessian of the reference state.

\section{Theory}

For two electronic excited states $|I \rangle$ and $|J \rangle$, we define the transition density matrix as:
\begin{equation}
    D^{IJ}_{pq} = \langle I |\hat{E}_{pq} | J \rangle ,
\end{equation}
where the spin-free transition operator is 
\begin{equation}
    \hat{E}_{pq} =\sum_\sigma \hat{a}^\dagger_{p \sigma} \hat{a}_{q\sigma}.
\end{equation}
We first assume that $I \neq J$ and $\omega_I \neq \omega_J$.
We introduce an arbitrary Hermitian one-body perturbation and give
\begin{equation}
    \hat{H}(\epsilon) =\hat{H}+\epsilon \hat{V},
\end{equation}
\begin{equation}
    \hat{V}= \sum_{pq} v_{pq} \hat{E}_{pq}.
\end{equation}
Then, we obtain
\begin{equation}
    \sum_{pq} v_{pq} D^{IJ}_{pq} = (\omega_J - \omega_I) \langle I | J ^\prime \rangle,
    \label{eq:tdm_perturbation_non_degenerate}
\end{equation}
and 
\begin{equation}
    \langle I| J ^\prime \rangle =\langle \Phi_0 | \hat{O}_I \hat{O}^{\dagger \prime}_J |\Phi_0 \rangle = \frac{1}{2} \langle \Phi_0 | [\hat{O}_I ,\hat{O}^{\dagger \prime}_J] - [\hat{O}^\prime_I ,\hat{O}^{\dagger}_J] |\Phi_0 \rangle .
    \label{eq:ij}
\end{equation}
Here, $X^\prime = \frac{dX}{d \epsilon}_{\epsilon=0}$. The first expression corresponds to the first EOM condition in our SA-TDDFT, and the second expression corresponds to the third EOM condition in our SA-TDDFT.

Based on our spin-adapted TDDFT (SA-TDDFT), we need to solve the following two equations:
\begin{equation}
    M R =\omega N R,
\end{equation}
and 
\begin{equation}
    L^\dagger M = \omega L^\dagger N.
\end{equation}

We now introduce an approximate expansion for the translation operator:
\begin{equation}
    \hat{O}_J^\dagger =\sum_{pq\Gamma \mu} R^J_{pq,\Gamma \mu} \hat{O}^\dagger_{pq}(\Gamma,\mu)
\end{equation}
and
\begin{equation}
    \hat{O}_I = \sum_{pq\Gamma \mu} (L^I_{pq,\Gamma \mu})^* \hat{O}_{pq}(\Gamma,\mu).
\end{equation}
The perturbed reference ROHF state is:
\begin{equation}
    |0 (\epsilon) \rangle = e^{\hat{\kappa}(\epsilon)} |0 \rangle
\end{equation}
where 
\begin{equation}
    \hat{\kappa}(\epsilon) = \sum_{rs} \kappa_{rs}(\epsilon) \hat{E}_{rs},
\end{equation}   
\begin{equation}
    \hat{\kappa}^\dagger =- \hat{\kappa},
\end{equation}
and 
\begin{equation}
    \hat{\kappa}(0)=0.
\end{equation}
The perturbed translation operator is 
\begin{equation}
    \hat{O}_{pq}(\Gamma,\mu;\epsilon) =  e^{\hat{\kappa}}  \hat{O}_{pq}(\Gamma,\mu) e^{-\hat{\kappa}}.
\end{equation}
By applying differentiation, we obtain
\begin{equation}
    \hat{O}_J^{\dagger \prime} = \sum_{pq\Gamma \mu} (R^J_{pq,\Gamma \mu})^\prime \hat{O}^\dagger_{pq}(\Gamma,\mu) +[\hat{\kappa}^\prime, \hat{O}^\dagger_J],
\end{equation}
and
\begin{equation}
    \hat{O}_I^\prime = \sum_{pq\Gamma \mu} (L^{I*}_{pq,\Gamma \mu})^\prime \hat{O}_{pq}(\Gamma,\mu) +[\hat{\kappa}^\prime, \hat{O}_I] .
\end{equation}
For the sake of convenience, we denote $\alpha =(pq, \Gamma \mu)$.
By applying substitution in Eq.~\ref{eq:ij}, we obtain
\begin{equation}
    \langle I | J^\prime \rangle = L_I^\dagger N R_J^\prime +\sum_{rs} \kappa_{rs}^\prime L_I^\dagger M_e^{rs} R_J,
\end{equation}
where 
\begin{equation}
    [M_e^{rs}]_{\alpha \beta} = \langle 0 |\{ \hat{O}_\alpha , \hat{E}_{rs}, \hat{O}^\dagger_\beta \} |0 \rangle
\end{equation}
and
\begin{equation}
    N_{\alpha \beta} = \langle 0 | \{ \hat{O}_\alpha , \hat{O}^\dagger_\beta \} |0 \rangle.
\end{equation}
Here, $\{ \}$ represents the EOM conditions in SA-TDDFT.

By differentiating $MR=\omega N R$, we obtain
\begin{equation}
    L^\dagger_I N R_J^\prime = \frac{L^\dagger_I M^\prime R_J}{\omega_J -\omega_I}.
    \label{eq:response_derivative_non_degenerate}
\end{equation}
By differentiating $M_{\alpha \beta}=\langle 0|\{ \hat{O}_\alpha, \hat{H}, \hat{O}^\dagger_\beta\}|0\rangle$, we obtain
\begin{equation}
    M^\prime = \sum_{rs} v_{rs} M_e^{rs} +\sum_{rs} \kappa^\prime_{rs} M_\kappa^{rs},
    \label{eq:M_derivative}
\end{equation}
where
\begin{equation}
    [M_\kappa^{rs}]_{\alpha \beta}= \langle 0 | \{ \hat{O}_\alpha, [\hat{H},\hat{E}_{rs}], \hat{O}^\dagger_\beta \} | 0 \rangle = \frac{\partial M_{\alpha \beta}(\kappa)}{\partial \kappa_{rs}}|_{\kappa=0}.
    \label{eq:mkappa}
\end{equation}
Thus,
\begin{equation}
    \sum_{rs} v_{rs} D^{IJ}_{rs} = \sum_{rs}v_{rs} L_I^\dagger M_e^{rs} R_J + \sum_{rs} \kappa^\prime_{rs} L^\dagger_I (M_\kappa^{rs}+(\omega_J-\omega_I)M_e^{rs}) R_J.
    \label{eq:tdm_contraction_non_degenerate}
\end{equation}
We define
\begin{equation}
    D_{rs}^{IJ,\mathrm{bil}} =\frac{1}{2}[L_I^\dagger M_e^{rs}R_J+(L_J^\dagger M_e^{sr}R_I)^*],
    \label{eq:tdm_bilinear}
\end{equation}
and 
\begin{equation}
    b^{IJ}_{rs} = \frac{1}{2} L^\dagger_I (M_\kappa^{rs}+(\omega_J-\omega_I)M_e^{rs})R_J - \frac{1}{2} [L^\dagger_J (M_\kappa^{sr}-(\omega_J-\omega_I)M_e^{sr})R_I ]^*.
    \label{eq:tdm_source}
\end{equation}
Then,
\begin{equation}
    \sum_{rs }  v_{rs}  D^{IJ}_{rs}  =\sum_{rs} v_{rs} D_{rs}^{IJ,\mathrm{bil}} +\sum_{rs} \kappa_{rs}^\prime b^{IJ}_{rs}. 
\end{equation}
We denote the reference state energy as
\begin{equation}
    E_0(\epsilon) = \langle 0| e^{-\hat{\kappa}} (\hat{H}+\epsilon \hat{V})e^{\hat{\kappa}} |0 \rangle.
\end{equation}
By using the BCH expansion, we obtain that
\begin{equation}
\begin{split}
  \mathcal A_{rs,pq} =\frac{\partial^2 E_0(\epsilon)}{\partial \kappa_{rs} \partial \kappa_{pq}}|_{\epsilon=0} =\frac{1}{2} \langle 0 | [[\hat{H},\hat{E}_{rs}],\hat{E}_{pq}]+[[\hat{H},\hat{E}_{pq}],\hat{E}_{rs}]| 0 \rangle.  
\end{split}
\end{equation}
For each $\epsilon$, the reference state satisfies
\begin{equation}
    \frac{\partial E_0(\epsilon)}{\partial \kappa_{rs}}=0.
\end{equation}
Differentiation by $\epsilon$ and setting $\epsilon=0$ gives
\begin{equation}
    \sum_{pq} \mathcal{A}_{rs,pq} \kappa^\prime_{pq} =-(n_s-n_r)v_{sr},
\end{equation}
where $n_p$ is the occupation number of the spatial orbital $p$.
We define
\begin{equation}
    \mathcal A^T t^{IJ} =- b^{IJ}.
\end{equation}
Thus, we obtain the important result in this work:
\begin{equation}
    D_{pq}^{IJ} = D_{pq}^{IJ,\mathrm{bil}} + D_{pq}^{IJ,\mathrm{rel}},
    \label{eq:tdm_decomposition}
\end{equation}
where 
\begin{equation}
    D_{pq}^{IJ,\mathrm{rel}} = (n_p-n_q) t_{qp}^{IJ}.
    \label{eq:tdm_relaxation}
\end{equation}
We now step into the derivation of explicit working equations. 
We define the following conventions:
\begin{equation}
    D_{PQ}= \langle \hat{a}^\dagger_P \hat{a}_Q \rangle,
\end{equation}
\begin{equation}
    f_{PQ} = \frac{\partial E}{\partial D_{PQ}},
\end{equation}
\begin{equation}
    K_{PQ,RS}= \frac{\partial^2 E}{\partial D_{PQ} \partial D_{SR}},
\end{equation}
and
\begin{equation}
    G_{PQ,RS,XY} =\frac{\partial^3 E}{\partial D_{PQ} \partial D_{SR} \partial D_{YX}}.
\end{equation}

Obviously,
\begin{equation}
\begin{aligned}
\mathcal A_{pq,rs}
={}&\frac12\sum_{\sigma}
\Big[
\delta_{qr}(n_p^\sigma+n_s^\sigma-2n_q^\sigma)f_{sp}^\sigma\\
&\qquad+
\delta_{ps}(n_q^\sigma+n_r^\sigma-2n_p^\sigma)f_{qr}^\sigma
\Big]\\
&+\sum_{\sigma\tau}
(n_q^\sigma-n_p^\sigma)
(n_s^\tau-n_r^\tau)
K_{qp,rs}^{\sigma\sigma,\tau\tau}.
\end{aligned}
\end{equation}
$M^{pq}_e$ has been formulated in SA-TDDFT, and the remaining term is $M_\kappa^{pq}$. Based on the second expression in Eq.~\ref{eq:mkappa}, we can differentiate the Fock and kernel in $M$ to obtain this term. It is noteworthy that we use the original formulation of $M$ to derive this without applying the Brillouin theorem before differentiation. We define 
\begin{equation}
\begin{aligned}
(f_\kappa^{rs})_{pq}^{\sigma}
={}&\delta_{qs}f_{pr}^{\sigma}
-\delta_{pr}f_{sq}^{\sigma}
\\
&+\sum_{\tau}
(n_s^{\tau}-n_r^{\tau})
K_{pq,rs}^{\sigma\sigma,\tau\tau},
\end{aligned}
\label{eq:fkappa}
\end{equation}
and
\begin{equation}
\begin{aligned}
(K_\kappa^{rs})_{pq,xy}^{\sigma\sigma',\tau\tau'}
={}&\delta_{qs}K_{pr,xy}^{\sigma\sigma',\tau\tau'}
-\delta_{pr}K_{sq,xy}^{\sigma\sigma',\tau\tau'}
\\
&+\delta_{xs}K_{pq,ry}^{\sigma\sigma',\tau\tau'}
-\delta_{yr}K_{pq,xs}^{\sigma\sigma',\tau\tau'}
\\
&+\sum_{\eta}
(n_s^{\eta}-n_r^{\eta})
G_{pq,xy,rs}^{\sigma\sigma',\tau\tau',\eta\eta}.
\end{aligned}
\label{eq:kkappa}
\end{equation}
By translating the Fock and kernel in $M$ to the aforementioned ones, we can obtain $M_\kappa^{pq}$. 

The following are the explicit formulations for the original $M$.
Every $M_{S_f}$ below is represented in exactly the same final response basis as $N_{S_f}$, including the same normalization, excitation--deexcitation pair reversal, phase convention, and zero-mode projection where required.

For $S_f=S_i+1$, the complete matrix in the ordered basis $(\mathrm{CV}(1)\mid\mathrm{VC}(1))$ is
\begin{equation}
M_{S_f}=\begin{pmatrix}
\delta_{ij}f_{ab}^{\alpha}-\delta_{ab}f_{ji}^{\beta}+K_{ai,bj}^{\mathrm{TC}}
&-K_{ai,jb}^{\mathrm{TC}}\\[3pt]
-K_{ia,bj}^{\mathrm{TC}}
&\delta_{ij}f_{ba}^{\beta}-\delta_{ab}f_{ij}^{\alpha}+K_{ia,jb}^{\mathrm{TC}}
\end{pmatrix}.
\end{equation}
For an open-shell reference, the $S_f=S_i$ matrix is written in the normalized order
\begin{equation}
\begin{split}
\bigl(&\mathrm{CV}(0),\mathrm{CO}(0),\mathrm{OV}(0),\mathrm{CV}(1)\\
&\mid\mathrm{VC}(0),\mathrm{OC}(0),\mathrm{VO}(0),\mathrm{VC}(1)\bigr).
\end{split}
\end{equation}
Its explicit working form is
\begin{equation}
M_{S_f}=\begin{pmatrix}
A & B\\[3pt]
B^* & \widetilde A^*
\end{pmatrix}.
\label{eq:M_equal}
\end{equation}
The detailed matrix elements are given in Table~\ref{tab:M_equal} in the Appendix.

For a closed-shell reference, the $S_f=S_i$ matrix must instead be derived directly in the order $(\mathrm{CV}(0)\mid\mathrm{VC}(0))$:
\begin{equation}
M_{S_f}=\begin{pmatrix}
\delta_{ij}f_{ab}-\delta_{ab}f_{ji}+K^S_{ai,bj}&K^S_{ai,jb}\\[3pt]
K^S_{ia,bj}&\delta_{ij}f_{ba}-\delta_{ab}f_{ij}+K^S_{ia,jb}
\end{pmatrix}.
\end{equation}
The open-shell expression contains $\sqrt{(S_i+1)/S_i}$ and therefore cannot be reduced to the closed-shell expression by simply substituting $S_i=0$.

For $S_f=S_i-1$, EOM1 gives
\begin{equation}
\widetilde M=
\frac{1}{S_i(2S_i-1)}M_+
-\frac{2S_i+1}{S_i(2S_i-1)}M_0
+\frac{2S_i+1}{2S_i-1}M_-.
\label{eq:M_lower_combination}
\end{equation}
The ordered basis is
\begin{equation}
\begin{split}
\bigl(&\mathrm{CV}(1),\mathrm{CO}(1),\mathrm{OV}(1),\mathrm{OO}(1),\mathrm{TT}(1)\\
&\mid\mathrm{VC}(1),\mathrm{OC}(1),\mathrm{VO}(1)\bigr),
\end{split}
\end{equation}
where OO contains the ordered pairs $t\ne u$ and TT contains the diagonal pairs $t=u$. The matrix is
\begingroup
\fontsize{9}{11}\selectfont
\setlength{\arraycolsep}{2.6pt}
\renewcommand{\arraystretch}{1.6}
\begin{equation}
M_{S_f}=\begin{pmatrix}
\widetilde A_{ai,bj}&\widetilde A_{ai,vj}&\widetilde A_{ai,bv}&\widetilde A_{ai,vw}&\widetilde L_{ai,vv}&\widetilde B_{ai,jb}&\widetilde B_{ai,jv}&\widetilde B_{ai,vb}\\
\widetilde A_{ui,bj}&\widetilde A_{ui,vj}&\widetilde A_{ui,bv}&\widetilde A_{ui,vw}&\widetilde L_{ui,vv}&\widetilde B_{ui,jb}&\widetilde B_{ui,jv}&\widetilde B_{ui,vb}\\
\widetilde A_{au,bj}&\widetilde A_{au,vj}&\widetilde A_{au,bv}&\widetilde A_{au,vw}&\widetilde L_{au,vv}&\widetilde B_{au,jb}&\widetilde B_{au,jv}&\widetilde B_{au,vb}\\
\widetilde A_{tu,bj}&\widetilde A_{tu,vj}&\widetilde A_{tu,bv}&\widetilde A_{tu,vw}&\widetilde L_{tu,vv}&\widetilde B_{tu,jb}&\widetilde B_{tu,jv}&\widetilde B_{tu,vb}\\
\widetilde L_{tt,bj}&\widetilde L_{tt,vj}&\widetilde L_{tt,bv}&\widetilde L_{tt,vw}&\widetilde D_{tt,vv}&\widetilde L_{tt,jb}&\widetilde L_{tt,jv}&\widetilde L_{tt,vb}\\
\widetilde B_{ia,bj}&\widetilde B_{ia,vj}&\widetilde B_{ia,bv}&\widetilde B_{ia,vw}&\widetilde L_{ia,vv}&\widetilde A_{ia,jb}&\widetilde A_{ia,jv}&\widetilde A_{ia,vb}\\
\widetilde B_{iu,bj}&\widetilde B_{iu,vj}&\widetilde B_{iu,bv}&\widetilde B_{iu,vw}&\widetilde L_{iu,vv}&\widetilde A_{iu,jb}&\widetilde A_{iu,jv}&\widetilde A_{iu,vb}\\
\widetilde B_{ua,bj}&\widetilde B_{ua,vj}&\widetilde B_{ua,bv}&\widetilde B_{ua,vw}&\widetilde L_{ua,vv}&\widetilde A_{ua,jb}&\widetilde A_{ua,jv}&\widetilde A_{ua,vb}
\end{pmatrix},
\label{eq:M_lower}
\end{equation}
\endgroup
where the last three block rows vanish. The entries in Eq.~\eqref{eq:M_lower} are obtained from Eq.~\eqref{eq:M_lower_combination} and the tables shown in the Appendix (Tables~\ref{tab:M_lower_A1}--\ref{tab:M_lower_OO}).

For the $S_f=S_i-1$ sector, the same TT zero-mode projection is applied simultaneously to $M_{S_f}$ and $N_{S_f}$, including every TT-connected block, so both matrices remain in the same final projected basis.

We now consider $I\ne J$ with $\omega_I=\omega_J=\omega_*$, and let $\mathcal D$ denote the isolated degenerate subspace. For $J\in\mathcal D$, we choose $|J(\epsilon)\rangle$ to be differentiable at $\epsilon=0$ and to satisfy
\begin{equation}
\begin{aligned}
\bigl[\hat H(\epsilon)-E_0(\epsilon)\bigr]|J(\epsilon)\rangle
&=\sum_{K\in\mathcal D}|K(\epsilon)\rangle\Omega_{KJ}(\epsilon),\\
\Omega_{IJ}(\epsilon)
&=\langle I(\epsilon)|\bigl[\hat H(\epsilon)-E_0(\epsilon)\bigr]|J(\epsilon)\rangle,\\
\langle I(\epsilon)|J(\epsilon)\rangle&=\delta_{IJ},
\qquad |J(0)\rangle=|J\rangle.
\end{aligned}
\label{eq:deg_block_states}
\end{equation}
Here $\Omega(0)=\omega_*\mathbf 1$, whereas $\Omega(\epsilon)$ need not be diagonal for $\epsilon\ne0$. Differentiating the first line and projecting onto $\langle I|$ at $\epsilon=0$, with $I\ne J$, gives
\begin{equation}
\begin{aligned}
\langle I|\hat V|J\rangle+\omega_*\langle I|J'\rangle
&=\omega_*\langle I|J'\rangle+\Omega'_{IJ},\\
\sum_{pq}v_{pq}D^{IJ}_{pq}&=\Omega'_{IJ}.
\end{aligned}
\label{eq:deg_transition}
\end{equation}
The second line replaces Eq.~\eqref{eq:tdm_perturbation_non_degenerate}. With the same operator expansion, the generalized eigenvalue equation becomes $MR_{\mathrm d}=NR_{\mathrm d}\Omega$, where $R_{\mathrm d}$ and $L_{\mathrm d}$ collect the corresponding right and left response vectors, with $L_{\mathrm d}^\dagger NR_{\mathrm d}=\mathbf 1$. Using $L_{\mathrm d}^\dagger M=\omega_*L_{\mathrm d}^\dagger N$ and $N'=0$ at $\epsilon=0$, differentiation yields
\begin{equation}
\begin{aligned}
L_{\mathrm d}^\dagger M'R_{\mathrm d}
+\omega_*L_{\mathrm d}^\dagger NR_{\mathrm d}'
&=\omega_*L_{\mathrm d}^\dagger NR_{\mathrm d}'+\Omega',\\
\Omega'_{IJ}&=L_I^\dagger M'R_J.
\end{aligned}
\label{eq:deg_response}
\end{equation}
The last line replaces Eq.~\eqref{eq:response_derivative_non_degenerate}, without division by $\omega_J-\omega_I$. Substituting the expression for $M'$ given in Eq.~\eqref{eq:M_derivative} into Eq.~\eqref{eq:deg_response}, and then using Eq.~\eqref{eq:deg_transition}, gives
\begin{equation}
\begin{aligned}
\sum_{rs}v_{rs}D^{IJ}_{rs}
={}&\sum_{rs}v_{rs}L_I^\dagger M_e^{rs}R_J\\
&+\sum_{rs}\kappa'_{rs}L_I^\dagger M_\kappa^{rs}R_J.
\end{aligned}
\label{eq:deg_contraction}
\end{equation}
This is the degenerate counterpart of Eq.~\eqref{eq:tdm_contraction_non_degenerate}. Applying the same Hermitianization, with $v_{rs}^*=v_{sr}$ and $(\kappa'_{rs})^*=-\kappa'_{sr}$, retains the bilinear density in Eq.~\eqref{eq:tdm_bilinear}, while Eq.~\eqref{eq:tdm_source} becomes
\begin{equation}
b^{IJ}_{rs}
=\frac12L_I^\dagger M_\kappa^{rs}R_J
-\frac12\left(L_J^\dagger M_\kappa^{sr}R_I\right)^*.
\label{eq:deg_source}
\end{equation}
The reference orbital-response equations and Eqs.~\eqref{eq:tdm_decomposition}--\eqref{eq:tdm_relaxation} retain their form with the source in Eq.~\eqref{eq:deg_source}, under the same regularity assumption on the reference orbital Hessian. 
The closed-shell TDHF reduction below has the same form for both the nondegenerate and degenerate cases.

In the closed-shell TDHF limit, a real RHF reference gives the singlet
eigenvalue equation
\begin{equation}
\begin{aligned}
&\sum_{jb}
\begin{pmatrix}
\delta_{ij}f_{ab}-\delta_{ab}f_{ji}+K^S_{ai,bj}&K^S_{ai,jb}\\[3pt]
K^S_{ia,bj}&\delta_{ij}f_{ba}-\delta_{ab}f_{ij}+K^S_{ia,jb}
\end{pmatrix}
\binom{X_{jb}^{K}}{Y_{jb}^{K}}\\
&\qquad=\omega_K\sum_{jb}
\begin{pmatrix}
\delta_{ij}\delta_{ab}&0\\
0&-\delta_{ij}\delta_{ab}
\end{pmatrix}
\binom{X_{jb}^{K}}{Y_{jb}^{K}}.
\end{aligned}
\label{eq:cs_eigen}
\end{equation}
The upper left and upper right blocks in Eq.~\eqref{eq:cs_eigen} are denoted
$A$ and $B$, respectively, with pair ordering $ai,bj$.
The transition density matrices are
\begin{equation}
\begin{aligned}
D_{ij}^{IJ,\mathrm{bil}}&=-\sum_a
(X_{ja}^I X_{ia}^J+Y_{ia}^I Y_{ja}^J),&
D_{ab}^{IJ,\mathrm{bil}}&=\sum_i
(X_{ia}^I X_{ib}^J+Y_{ib}^I Y_{ia}^J),\\
D_{ai}^{IJ,\mathrm{bil}}&=D_{ia}^{IJ,\mathrm{bil}}=0,&
D_{ij}^{IJ,\mathrm{rel}}&=D_{ab}^{IJ,\mathrm{rel}}=0,
\end{aligned}
\label{eq:cs_bil}
\end{equation}
\begin{equation}
\binom{D_{ai}^{IJ,\mathrm{rel}}}{D_{ia}^{IJ,\mathrm{rel}}}
=-\begin{pmatrix}A&B\\B&A\end{pmatrix}^{-1}
\binom{b_{ai}^{IJ}}{b_{ai}^{JI}},
\label{eq:cs_present}
\end{equation}
where
\begin{equation}
\begin{aligned}
b_{ai}^{IJ}={}&\sum_{pq}D_{pq}^{IJ,\mathrm{bil}}
(f_\kappa^{ai})_{pq}\\
&+\sum_{jb,kc}\Big[
X_{jb}^{I}X_{kc}^{J}(K_\kappa^{ai})^S_{bj,ck}
+X_{jb}^{I}Y_{kc}^{J}(K_\kappa^{ai})^S_{bj,kc}\\
&\hspace{8mm}
+Y_{jb}^{I}X_{kc}^{J}(K_\kappa^{ai})^S_{jb,ck}
+Y_{jb}^{I}Y_{kc}^{J}(K_\kappa^{ai})^S_{jb,kc}\Big],
\end{aligned}
\label{eq:cs_source}
\end{equation}
with
\begin{equation}
\begin{aligned}
(f_\kappa^{ai})_{pq}
&=\delta_{qi}f_{pa}-\delta_{pa}f_{iq}+K^S_{pq,ai},\\
(K_\kappa^{ai})^S_{pq,xy}
&=\delta_{qi}K^S_{pa,xy}-\delta_{pa}K^S_{iq,xy}
+\delta_{xi}K^S_{pq,ay}-\delta_{ya}K^S_{pq,xi}.
\end{aligned}
\label{eq:cs_derivatives}
\end{equation}

The following expressions use the notation and density matrix convention
of this work. Li and Liu's EOM density~\cite{LiLiu2014}, Eq.~(31), is
\begin{equation}
D_{pq}^{IJ,\mathrm{EOM}}=D_{pq}^{IJ,\mathrm{bil}}.
\label{eq:cs_EOM}
\end{equation}
Their quadratic response (QR) density, Eqs.~(112) and (126), has the mixed blocks
\begin{equation}
\binom{D_{ai}^{IJ,\mathrm{QR}}}{D_{ia}^{IJ,\mathrm{QR}}}
=-\begin{pmatrix}
A+\Delta\mathbf{1}&B\\B&A-\Delta\mathbf{1}
\end{pmatrix}^{-1}
\binom{b_{ai}^{IJ}}{b_{ai}^{JI}}, \Delta=\omega_J-\omega_I.
\label{eq:cs_QR}
\end{equation}
Alguire et al.~\cite{Alguire2015}, Eq.~(72), and
Ou et al.~\cite{Ou2015PW}, Eq.~(86), give the same pseudo-wavefunction (PW) density,
\begin{equation}
D_{ai}^{IJ,\mathrm{PW}}=D_{ia}^{IJ,\mathrm{PW}}
=-\frac12\sum_{bj}[(A+B)^{-1}]_{ai,bj}
(b_{bj}^{IJ}+b_{bj}^{JI}).
\label{eq:cs_PW}
\end{equation}

The $ij$ and $ab$ blocks are identical in all four formulations.
EOM sets both mixed blocks to zero. The present and QR densities use the
same $b$, while QR adds $+\Delta\mathbf{1}$ and $-\Delta\mathbf{1}$ to
the two diagonal response blocks. The two PW formulas are identical and
satisfy $D_{ai}^{IJ,\mathrm{PW}}=D_{ia}^{IJ,\mathrm{PW}}
=\tfrac12(D_{ai}^{IJ,\mathrm{rel}}+D_{ia}^{IJ,\mathrm{rel}})$. For the transition electric dipole moment, PW gives the same result as our method.

For electric dipole transitions between states $I$ and $J$ of the same
total spin, the transition moment is
\begin{equation}
\boldsymbol\mu_{IJ}=-\sum_{pq}\mathbf r_{pq}D_{pq}^{IJ}
=\boldsymbol\mu_{IJ}^{\mathrm{bil}}+\boldsymbol\mu_{IJ}^{\mathrm{rel}},
\qquad \mathbf r_{pq}=\langle p|\mathbf r|q\rangle.
\label{eq:transition_dipole}
\end{equation}
For $\Delta=\omega_J-\omega_I>0$, the isotropic oscillator strength in
the length representation is
\begin{equation}
f_{IJ}=\frac{2}{3}\Delta\,|\boldsymbol\mu_{IJ}|^2.
\label{eq:oscillator_strength}
\end{equation}

\section{Results and discussion}
\label{sec:results}

The calculations consider CO and para-benzyne using the BHHLYP functional
and the aug-cc-pVTZ basis set. The spin-adapted calculations employ the
Tamm--Dancoff approximation (SA-TDA), with six target roots for each
system. CO is treated with a closed-shell singlet reference in the
$S_f=S_i=0$ sector, whereas para-benzyne uses a triplet restricted
open-shell Kohn--Sham reference and singlet target states in the
$S_f=S_i-1$ sector. The CO bond length is $1.128$~\AA; the complete
geometries and calculation settings for both systems are specified in
the supporting information.

The wavefunction reference is state-averaged complete active space
self-consistent field (SA-CASSCF), using ten electrons in nine orbitals
for CO and eight electrons in eight orbitals for para-benzyne. Seven
singlet states are included with equal weights in each orbital
optimization, counting the individual components of degenerate states
separately. All states selected for the transition-property comparisons
participate in the state average. The CASSCF calculations use density
fitting, whereas the SA-TDA calculations do not. SA-TDA and CASSCF are
evaluated at the same fixed geometries and in the same orbital basis
set. Hereafter, CAS denotes this finite-active-space SA-CASSCF reference.

We compare the unrelaxed density
$D^{IJ,\mathrm{bil}}$ (U) with the full density
$D^{IJ,\mathrm{bil}}+D^{IJ,\mathrm{rel}}$ (R).
The U and R calculations use identical SA-TDA excitation energies and
response vectors; their differences in transition properties arise
solely from the density correction. For transitions between electronic
terms, the reported effective transition dipole is defined by
\begin{equation*}
 \mu_{\mathrm{eff}}^2
 =\frac{1}{g_I}\sum_{a\in I}\sum_{b\in J}
   \left|\boldsymbol{\mu}_{ab}\right|^2,
 \qquad
 f_{IJ}=\frac{2}{3}\Delta E_{IJ}\mu_{\mathrm{eff}}^2,
\end{equation*}
where $I$ is the lower-energy term, $g_I$ is its degeneracy, and its
components are assumed to be equally populated. The transition
strength is summed over all final components and averaged over the
initial components. For nondegenerate states,
$\mu_{\mathrm{eff}}=|\boldsymbol{\mu}_{IJ}|$. Both components of the
CO ${}^{1}\Pi$ and ${}^{1}\Delta$ terms are included. 

Table~\ref{tab:co_results} compares the energy gaps, effective transition
dipoles, and oscillator strengths between the ${}^{1}\Pi$,
${}^{1}\Sigma^{-}$, ${}^{1}\Delta$, and excited ${}^{1}\Sigma^{+}$ terms of CO. The SA-TDA gaps retain the CASSCF
ordering of these terms, but their accuracy is not uniform. The
${}^{1}\Pi\to{}^{1}\Delta$ gap differs from CASSCF by only
$0.0316$~eV, whereas the ${}^{1}\Pi\to{}^{1}\Sigma^{-}$ gap is
underestimated by $0.2650$~eV and the
${}^{1}\Pi\to{}^{1}\Sigma^{+}$ gap is overestimated by $0.8318$~eV.
These differences are relevant when comparing oscillator strengths,
which depend on both the energy gap and the transition density.

\begin{table}[htbp]
\centering
\caption{Interstate transition properties of CO at
$r_{\mathrm{CO}}=1.128$~\AA. $\Delta E$ is in eV,
$\mu=\mu_{\mathrm{eff}}$ in atomic units, and $f$ is dimensionless.
CAS: SA7-CASSCF(10e,9o); SA: BHHLYP SA-TDA.
$\approx0$ indicates negligible values at the reported precision.}
\label{tab:co_results}
\begingroup
\fontsize{10.5}{12.5}\selectfont
\setlength{\tabcolsep}{2.5pt}
\renewcommand{\arraystretch}{1.3}
\begin{tabular}{@{}lrrrrrrrr@{}}
\toprule
Transition & $\Delta E_{\rm CAS}$ & $\Delta E_{\rm SA}$
& $\mu_{\rm CAS}$ & $\mu_{\rm U}$ & $\mu_{\rm R}$
& $f_{\rm CAS}$ & $f_{\rm U}$ & $f_{\rm R}$\\
\midrule
${}^{1}\Pi\to{}^{1}\Sigma^{-}$
& 1.1671 & 0.9021 & 0.2057 & 0.2543 & 0.2168
& 0.001210 & 0.001429 & 0.001039\\
${}^{1}\Pi\to{}^{1}\Delta$
& 1.2274 & 1.2590 & 0.1901 & 0.3184 & 0.2158
& 0.001087 & 0.003126 & 0.001436\\
${}^{1}\Pi\to{}^{1}\Sigma^{+}$
& 1.6825 & 2.5143 & 0.2943 & 0.5888 & 0.5439
& 0.003570 & 0.021353 & 0.018223\\
${}^{1}\Sigma^{-}\to{}^{1}\Delta$
& 0.0603 & 0.3568 & $\approx0$ & $\approx0$ & $\approx0$
& $\approx0$ & $\approx0$ & $\approx0$\\
${}^{1}\Sigma^{-}\to{}^{1}\Sigma^{+}$
& 0.5154 & 1.6122 & $\approx0$ & $\approx0$ & $\approx0$
& $\approx0$ & $\approx0$ & $\approx0$\\
${}^{1}\Delta\to{}^{1}\Sigma^{+}$
& 0.4551 & 1.2554 & $\approx0$ & $\approx0$ & $\approx0$
& $\approx0$ & $\approx0$ & $\approx0$\\
\bottomrule
\end{tabular}
\endgroup
\end{table}

Including orbital relaxation reduces all three nonzero transition
dipoles and brings them closer to the CASSCF values. For
${}^{1}\Pi\to{}^{1}\Sigma^{-}$, $\mu$ decreases from $0.2543$ to
$0.2168$~a.u., compared with $0.2057$~a.u.\ from CASSCF; the relative
deviation decreases from $23.6\%$ to $5.4\%$. The largest relative
change occurs for ${}^{1}\Pi\to{}^{1}\Delta$, where $\mu$ decreases
from $0.3184$ to $0.2158$~a.u., compared with $0.1901$~a.u.\ from
CASSCF. The corresponding relative deviation decreases from
$67.5\%$ to $13.5\%$. Since the energy gap for this transition is
already close to the CASSCF value, it provides a direct illustration
of the importance of the transition-density correction.

For ${}^{1}\Pi\to{}^{1}\Delta$, the oscillator strength decreases
from $0.003126$ to $0.001436$, approaching the CASSCF value of
$0.001087$. This is a $54.1\%$ reduction relative to the unrelaxed
result. For ${}^{1}\Pi\to{}^{1}\Sigma^{-}$, $f$ changes from
$0.001429$ to $0.001039$, compared with $0.001210$ from CASSCF.
Although the relaxed dipole remains slightly larger than the CASSCF
value, the relaxed oscillator strength is smaller because the SA-TDA
energy gap is underestimated. The unrelaxed oscillator strength
benefits from partial cancellation between its overestimated dipole
and underestimated gap. Consequently, agreement in $f$ alone can
conceal a larger discrepancy in the transition density.

The ${}^{1}\Pi\to{}^{1}\Sigma^{+}$ transition remains less
accurate. Relaxation reduces $\mu$ from $0.5888$ to $0.5439$~a.u.,
but the latter is still $84.8\%$ above the CASSCF value of
$0.2943$~a.u. The oscillator strength decreases from $0.021353$ to
$0.018223$, which remains about $5.10$ times the CASSCF value.
Both the larger dipole and the overestimated energy gap contribute
to this residual discrepancy. Thus, the relaxation correction
improves this transition but does not by itself ensure quantitative
agreement with the wavefunction reference. The remaining three
transitions have negligible dipoles and oscillator strengths at
the reported precision in all three treatments; adding relaxation
does not produce an appreciable intensity for these transitions.

Para-benzyne tests the singlet target states obtained from an open-shell
triplet reference. Table~\ref{tab:para_results} compares transitions
between its ${}^{1}A_g$, ${}^{1}B_{3g}$, ${}^{1}B_{2g}$, and
${}^{1}B_{1u}$ states. SA-TDA retains their CASSCF ordering but
underestimates the three energy gaps from ${}^{1}A_g$ by
$0.9629$, $0.5276$, and $0.6990$~eV, respectively. The
${}^{1}A_g\to{}^{1}B_{3g}$,
${}^{1}A_g\to{}^{1}B_{2g}$, and
${}^{1}B_{3g}\to{}^{1}B_{2g}$ transitions have negligible
dipoles and oscillator strengths at the reported precision in
all three treatments.

\begin{table}[htbp]
\centering
\caption{Interstate transition properties of para-benzyne with
aug-cc-pVTZ. CAS denotes SA7-CASSCF(8e,8o), and SA denotes BHHLYP
SA-TDA with a triplet reference and singlet target states.
Units and the definitions of U and R are the same as in
Table~\ref{tab:co_results}; $\approx0$ indicates negligible values
at the reported precision.}
\label{tab:para_results}
\begingroup
\fontsize{10.5}{12.5}\selectfont
\setlength{\tabcolsep}{2.5pt}
\renewcommand{\arraystretch}{1.3}
\begin{tabular}{@{}lrrrrrrrr@{}}
\toprule
Transition & $\Delta E_{\rm CAS}$ & $\Delta E_{\rm SA}$
& $\mu_{\rm CAS}$ & $\mu_{\rm U}$ & $\mu_{\rm R}$
& $f_{\rm CAS}$ & $f_{\rm U}$ & $f_{\rm R}$\\
\midrule
${}^{1}A_g\to{}^{1}B_{3g}$
& 3.6006 & 2.6377 & $\approx0$ & $\approx0$ & $\approx0$
& $\approx0$ & $\approx0$ & $\approx0$\\
${}^{1}A_g\to{}^{1}B_{2g}$
& 3.8775 & 3.3499 & $\approx0$ & $\approx0$ & $\approx0$
& $\approx0$ & $\approx0$ & $\approx0$\\
${}^{1}A_g\to{}^{1}B_{1u}$
& 4.2265 & 3.5275 & 0.1701 & 0.1350 & 0.0860
& 0.002996 & 0.001574 & 0.000639\\
${}^{1}B_{3g}\to{}^{1}B_{2g}$
& 0.2769 & 0.7122 & $\approx0$ & $\approx0$ & $\approx0$
& $\approx0$ & $\approx0$ & $\approx0$\\
${}^{1}B_{3g}\to{}^{1}B_{1u}$
& 0.6259 & 0.8898 & 1.0862 & 1.7352 & 1.0866
& 0.018093 & 0.065635 & 0.025740\\
${}^{1}B_{2g}\to{}^{1}B_{1u}$
& 0.3490 & 0.1776 & 0.3608 & 0.6053 & 0.3282
& 0.001113 & 0.001594 & 0.000469\\
\bottomrule
\end{tabular}
\endgroup
\end{table}

The largest nonzero transition dipole is found for
${}^{1}B_{3g}\to{}^{1}B_{1u}$. Orbital relaxation reduces
$\mu$ from $1.7352$ to $1.0866$~a.u., in close agreement with
the CASSCF value of $1.0862$~a.u. The relative deviation decreases
from $59.7\%$ to approximately $0.04\%$. The oscillator strength
decreases from $0.065635$ to $0.025740$, compared with
$0.018093$ from CASSCF. Its remaining $42.3\%$ overestimate
is almost entirely accounted for by the energy gap:
$0.8898$~eV from SA-TDA versus $0.6259$~eV from CASSCF,
a difference of $42.2\%$. This transition therefore illustrates
that a nearly matching transition dipole can still yield a
substantially different oscillator strength.

For ${}^{1}B_{2g}\to{}^{1}B_{1u}$, $\mu$ decreases from
$0.6053$ to $0.3282$~a.u., compared with $0.3608$~a.u.\
from CASSCF. The absolute relative deviation decreases from
$67.8\%$ to $9.0\%$. Nevertheless, $f$ changes from
$0.001594$ to $0.000469$, compared with $0.001113$ from
CASSCF, so its absolute relative deviation increases from
$43.2\%$ to $57.9\%$. The SA-TDA gap, $0.1776$~eV,
is about half the CASSCF gap of $0.3490$~eV. In the
unrelaxed result, this underestimated gap partly compensates
for the overestimated dipole. Reducing the dipole improves
its agreement with CASSCF while exposing the energy-gap
error in the oscillator strength.

The ${}^{1}A_g\to{}^{1}B_{1u}$ transition shows a different
trend. Its unrelaxed dipole of $0.1350$~a.u.\ is already
below the CASSCF value of $0.1701$~a.u.; relaxation lowers
it further to $0.0860$~a.u., increasing the relative
underestimate from $20.6\%$ to $49.4\%$. The oscillator
strength likewise decreases from $0.001574$ to $0.000639$,
moving further below the CASSCF value of $0.002996$.
Thus, although relaxation brings the other two nonzero
para-benzyne dipoles closer to CASSCF, its effect on agreement
with the wavefunction reference is transition dependent.

\section{Conclusions}
\label{sec:conclusions}

We have formulated interstate one-particle transition density matrices
within SA-TDDFT by retaining the response of both the reference orbitals
and the excitation operators in the generalized EOM eigenvalue problem.
The resulting density separates into a bilinear contribution and an
orbital relaxation contribution obtained from the reference orbital
Hessian. Explicit working equations are given for the spin-adapted
response sectors, together with a treatment of degenerate states.
Contraction of the resulting density with one-electron integrals
provides interstate transition moments.

The CO and para-benzyne calculations show that orbital relaxation can
substantially affect transition dipoles for both closed-shell and
open-shell references within the Tamm--Dancoff approximation.
It brings five of the six nonzero dipoles closer to the SA-CASSCF
values, with particularly close agreement for the
${}^{1}B_{3g}\to{}^{1}B_{1u}$ transition of para-benzyne.
However, the ${}^{1}A_g\to{}^{1}B_{1u}$ transition exhibits
a larger discrepancy after relaxation, demonstrating that the
improvement is not uniform. Oscillator strengths also depend on
the SA-TDA energy gaps, and closer agreement with CASSCF in the
dipole can coexist with a larger deviation in $f$ when cancellation
between the dipole and energy-gap errors is reduced. These results
support explicit evaluation
of orbital relaxation and separate assessment of transition
dipoles and energy gaps when interpreting interstate optical
intensities.

\suppdata{The supplementary archive contains \texttt{si.pdf} and \texttt{tdm.py}, the Python calculation input for the CO and para-benzyne results reported in this work.}

\section*{Data and software availability}
Data is fully public. Our software IQC (v2.1.0) \cite{Zhang2026} can be accessed through our github release. \cite{Zhang2026IQC}

\Needspace{10\baselineskip}
\appendix
\section{Matrix elements for \texorpdfstring{$M$}{M}}
\label{app:M_elements}
\begingroup
\setstretch{1}
\fontsize{9.5}{12}\selectfont
\setlength{\tabcolsep}{4pt}
\setlength{\LTcapwidth}{\textwidth}
\begin{longtable}{@{}p{.25\textwidth}p{.73\textwidth}@{}}
\caption{Matrix Elements of Eq.~\eqref{eq:M_equal}. Only the upper triangular parts of $A$ and $B$ as well as the CV(1)--CV(1) block of $\widetilde A$ are listed since $A$ is Hermitian, $B$ is symmetric, and $\widetilde A$ differs from $A$ only in the triplet--triplet block.}\label{tab:M_equal}\\
\toprule
Blocks & Matrix elements\\
\midrule
\endfirsthead
\toprule
Blocks & Matrix elements\\
\midrule
\endhead
\midrule
\endfoot
\bottomrule
\endlastfoot
CV(0)--CV(0) & $\displaystyle\begin{aligned}[t][A]_{ai,bj}&=\frac{\delta_{ij} \left(f_{ab}^{\alpha} + f_{ab}^{\beta}\right)}{2} - \frac{\delta_{ab} \left(f_{ji}^{\alpha} + f_{ji}^{\beta}\right)}{2} +K_{ai,bj}^{S}\end{aligned}$\\[9pt]
CV(0)--CO(0) & $\displaystyle\begin{aligned}[t][A]_{ai,vj}&=\frac{\sqrt{2} \delta_{ij} \left(f_{av}^{\alpha} + 2 f_{av}^{\beta}\right)}{4}\\[1pt]&\quad +\frac{\sqrt{2} \left(K_{ai,vj}^{\alpha\alpha,\beta\beta} + K_{ai,vj}^{\beta\beta,\beta\beta}\right)}{2}\end{aligned}$\\[9pt]
CV(0)--OV(0) & $\displaystyle\begin{aligned}[t][A]_{ai,bv}&=- \frac{\sqrt{2} \delta_{ab} \left(2 f_{vi}^{\alpha} + f_{vi}^{\beta}\right)}{4}\\[1pt]&\quad +\frac{\sqrt{2} \left(K_{ai,bv}^{\alpha\alpha,\alpha\alpha} + K_{ai,bv}^{\beta\beta,\alpha\alpha}\right)}{2}\end{aligned}$\\[9pt]
CO(0)--CO(0) & $\displaystyle\begin{aligned}[t][A]_{ui,vj}&=\delta_{ij} f_{uv}^{\beta} - \delta_{uv} f_{ji}^{\beta} +K_{ui,vj}^{\beta\beta,\beta\beta}\end{aligned}$\\[9pt]
CO(0)--OV(0) & $\displaystyle\begin{aligned}[t][A]_{ui,bv}&=K_{ui,bv}^{\beta\beta,\alpha\alpha}\end{aligned}$\\[9pt]
OV(0)--OV(0) & $\displaystyle\begin{aligned}[t][A]_{au,bv}&=\delta_{uv} f_{ab}^{\alpha} - \delta_{ab} f_{vu}^{\alpha} +K_{au,bv}^{\alpha\alpha,\alpha\alpha}\end{aligned}$\\[9pt]
CV(0)--CV(1) & $\displaystyle\begin{aligned}[t][A]_{ai,bj}&=\frac{\delta_{ij} \sqrt{\tfrac{S_i+1}{S_i}} \left(- f_{ab}^{\alpha} + f_{ab}^{\beta}\right)}{2}\\[1pt]&\quad +\frac{\delta_{ab} \sqrt{\tfrac{S_i+1}{S_i}} \left(f_{ji}^{\alpha} - f_{ji}^{\beta}\right)}{2}\\[1pt]&\quad - K_{ai,bj}^{P} \sqrt{\tfrac{S_i+1}{S_i}}\end{aligned}$\\[9pt]
CO(0)--CV(1) & $\displaystyle\begin{aligned}[t][A]_{ui,bj}&=\frac{\sqrt{2} \delta_{ij} \sqrt{\tfrac{S_i+1}{S_i}} \left(- f_{ub}^{\alpha} + 2 f_{ub}^{\beta}\right)}{4}\\[1pt]&\quad +\frac{\sqrt{2} \sqrt{\tfrac{S_i+1}{S_i}} \left(- K_{ui,bj}^{\beta\beta,\alpha\alpha} + K_{ui,bj}^{\beta\beta,\beta\beta}\right)}{2}\end{aligned}$\\[9pt]
OV(0)--CV(1) & $\displaystyle\begin{aligned}[t][A]_{au,bj}&=\frac{\sqrt{2} \delta_{ab} \sqrt{\tfrac{S_i+1}{S_i}} \left(2 f_{ju}^{\alpha} - f_{ju}^{\beta}\right)}{4}\\[1pt]&\quad +\frac{\sqrt{2} \sqrt{\tfrac{S_i+1}{S_i}} \left(- K_{au,bj}^{\alpha\alpha,\alpha\alpha} + K_{au,bj}^{\alpha\alpha,\beta\beta}\right)}{2}\end{aligned}$\\[9pt]
CV(1)--CV(1) & $\displaystyle\begin{aligned}[t][A]_{ai,bj}&=\frac{\delta_{ij} \left(f_{ab}^{\alpha} + f_{ab}^{\beta} - \frac{f_{ab}^{\alpha}}{S_i} + \frac{f_{ab}^{\beta}}{S_i}\right)}{2}\\[1pt]&\quad +\frac{\delta_{ab} \left(- f_{ji}^{\alpha} - f_{ji}^{\beta} - \frac{f_{ji}^{\alpha}}{S_i} + \frac{f_{ji}^{\beta}}{S_i}\right)}{2} +K_{ai,bj}^{\mathrm{TC}}\end{aligned}$\\[9pt]

CV(0)--VC(0) & $\displaystyle\begin{aligned}[t][B]_{ai,jb}&=K_{ai,jb}^{S}\end{aligned}$\\[9pt]
CV(0)--OC(0) & $\displaystyle\begin{aligned}[t][B]_{ai,jv}&=\frac{\sqrt{2} \left(K_{ai,jv}^{\alpha\alpha,\beta\beta} + K_{ai,jv}^{\beta\beta,\beta\beta}\right)}{2}\end{aligned}$\\[9pt]
CV(0)--VO(0) & $\displaystyle\begin{aligned}[t][B]_{ai,vb}&=\frac{\sqrt{2} \left(K_{ai,vb}^{\alpha\alpha,\alpha\alpha} + K_{ai,vb}^{\beta\beta,\alpha\alpha}\right)}{2}\end{aligned}$\\[9pt]
CO(0)--OC(0) & $\displaystyle\begin{aligned}[t][B]_{ui,jv}&=K_{ui,jv}^{\beta\beta,\beta\beta}\end{aligned}$\\[9pt]
CO(0)--VO(0) & $\displaystyle\begin{aligned}[t][B]_{ui,vb}&=\frac{\delta_{uv} \left(- f_{bi}^{\alpha} + f_{bi}^{\beta}\right)}{2} +K_{ui,vb}^{\beta\beta,\alpha\alpha}\end{aligned}$\\[9pt]
OV(0)--VO(0) & $\displaystyle\begin{aligned}[t][B]_{au,vb}&=K_{au,vb}^{\alpha\alpha,\alpha\alpha}\end{aligned}$\\[9pt]
CV(0)--VC(1) & $\displaystyle\begin{aligned}[t][B]_{ai,jb}&=- K_{ai,jb}^{P} \sqrt{\tfrac{S_i+1}{S_i}}\end{aligned}$\\[9pt]
CO(0)--VC(1) & $\displaystyle\begin{aligned}[t][B]_{ui,jb}&=\frac{\sqrt{2} \sqrt{\tfrac{S_i+1}{S_i}} \left(- K_{ui,jb}^{\beta\beta,\alpha\alpha} + K_{ui,jb}^{\beta\beta,\beta\beta}\right)}{2}\end{aligned}$\\[9pt]
OV(0)--VC(1) & $\displaystyle\begin{aligned}[t][B]_{au,jb}&=\frac{\sqrt{2} \sqrt{\tfrac{S_i+1}{S_i}} \left(- K_{au,jb}^{\alpha\alpha,\alpha\alpha} + K_{au,jb}^{\alpha\alpha,\beta\beta}\right)}{2}\end{aligned}$\\[9pt]
CV(1)--VC(1) & $\displaystyle\begin{aligned}[t][B]_{ai,jb}&=K_{ai,jb}^{\mathrm{TC}}\end{aligned}$\\[9pt]
VC(1)--VC(1) & $\displaystyle\begin{aligned}[t][\widetilde A]_{ai,bj}&=\frac{\delta_{ij} \left(f_{ab}^{\alpha} + f_{ab}^{\beta} + \frac{f_{ab}^{\alpha}}{S_i} - \frac{f_{ab}^{\beta}}{S_i}\right)}{2}\\[1pt]&\quad +\frac{\delta_{ab} \left(- f_{ji}^{\alpha} - f_{ji}^{\beta} + \frac{f_{ji}^{\alpha}}{S_i} - \frac{f_{ji}^{\beta}}{S_i}\right)}{2} +K_{ai,bj}^{\mathrm{TC}}\end{aligned}$\\[9pt]
\end{longtable}
\endgroup

\begingroup
\setstretch{1}
\fontsize{9.5}{12}\selectfont
\setlength{\tabcolsep}{6pt}
\setlength{\LTcapwidth}{\textwidth}
\begin{longtable}{@{}p{.15\textwidth}p{.82\textwidth}@{}}
\caption{Matrix Elements of $M_+$, $M_0$, and $M_-$ without OO Transitions and Crossing Blocks, Based on EOM1}\label{tab:M_lower_A1}\\
\toprule
Blocks & Matrix elements\\
\midrule
\endfirsthead
\toprule
Blocks & Matrix elements\\
\midrule
\endhead
\midrule
\endfoot
\bottomrule
\endlastfoot
CV--CV & $\displaystyle\begin{aligned}[t][M_{+}]_{ai,bj}&=K_{ai,bj}^{\mathrm{TC}} - \delta_{ab} f_{ji}^{\beta} + \delta_{ij} f_{ab}^{\alpha}\\[2pt][M_{0}]_{ai,bj}&=K_{ai,bj}^{\mathrm{TC}} - \frac{\delta_{ab} f_{ji}^{\alpha}}{2} - \frac{\delta_{ab} f_{ji}^{\beta}}{2} + \frac{\delta_{ij} f_{ab}^{\alpha}}{2} + \frac{\delta_{ij} f_{ab}^{\beta}}{2}\\[2pt][M_{-}]_{ai,bj}&=K_{ai,bj}^{\mathrm{TC}} - \delta_{ab} f_{ji}^{\alpha} + \delta_{ij} f_{ab}^{\beta}\end{aligned}$\\[12pt]
VC--VC & $\displaystyle\begin{aligned}[t][M_{+}]_{ia,jb}&=0\\[2pt][M_{0}]_{ia,jb}&=0\\[2pt][M_{-}]_{ia,jb}&=0\end{aligned}$\\[12pt]
CV--CO & $\displaystyle\begin{aligned}[t][M_{+}]_{ai,vj}&=\delta_{ij} f_{av}^{\alpha}\\[2pt][M_{0}]_{ai,vj}&=\frac{- K_{ai,vj}^{\alpha\alpha,\beta\beta} + K_{ai,vj}^{\beta\beta,\beta\beta} + \delta_{ij} f_{av}^{\alpha} + \delta_{ij} f_{av}^{\beta}}{2}\\[2pt][M_{-}]_{ai,vj}&=K_{ai,vj}^{\mathrm{TC}} + \delta_{ij} f_{av}^{\beta}\end{aligned}$\\[12pt]
VC--OC & $\displaystyle\begin{aligned}[t][M_{+}]_{ia,jv}&=0\\[2pt][M_{0}]_{ia,jv}&=0\\[2pt][M_{-}]_{ia,jv}&=0\end{aligned}$\\[12pt]
CO--CV & $\displaystyle\begin{aligned}[t][M_{+}]_{ui,bj}&=0\\[2pt][M_{0}]_{ui,bj}&=\frac{- K_{ui,bj}^{\beta\beta,\alpha\alpha} + K_{ui,bj}^{\beta\beta,\beta\beta} + \delta_{ij} f_{ub}^{\beta}}{2}\\[2pt][M_{-}]_{ui,bj}&=K_{ui,bj}^{\mathrm{TC}} + \delta_{ij} f_{ub}^{\beta}\end{aligned}$\\[12pt]
OC--VC & $\displaystyle\begin{aligned}[t][M_{+}]_{iu,jb}&=0\\[2pt][M_{0}]_{iu,jb}&=0\\[2pt][M_{-}]_{iu,jb}&=0\end{aligned}$\\[12pt]
CV--OV & $\displaystyle\begin{aligned}[t][M_{+}]_{ai,bv}&=- \delta_{ab} f_{vi}^{\beta}\\[2pt][M_{0}]_{ai,bv}&=\frac{K_{ai,bv}^{\alpha\alpha,\alpha\alpha} - K_{ai,bv}^{\beta\beta,\alpha\alpha} - \delta_{ab} f_{vi}^{\alpha} - \delta_{ab} f_{vi}^{\beta}}{2}\\[2pt][M_{-}]_{ai,bv}&=K_{ai,bv}^{\mathrm{TC}} - \delta_{ab} f_{vi}^{\alpha}\end{aligned}$\\[12pt]
VC--VO & $\displaystyle\begin{aligned}[t][M_{+}]_{ia,vb}&=0\\[2pt][M_{0}]_{ia,vb}&=0\\[2pt][M_{-}]_{ia,vb}&=0\end{aligned}$\\[12pt]
OV--CV & $\displaystyle\begin{aligned}[t][M_{+}]_{au,bj}&=0\\[2pt][M_{0}]_{au,bj}&=\frac{K_{au,bj}^{\alpha\alpha,\alpha\alpha} - K_{au,bj}^{\alpha\alpha,\beta\beta} - \delta_{ab} f_{ju}^{\alpha}}{2}\\[2pt][M_{-}]_{au,bj}&=K_{au,bj}^{\mathrm{TC}} - \delta_{ab} f_{ju}^{\alpha}\end{aligned}$\\[12pt]
VO--VC & $\displaystyle\begin{aligned}[t][M_{+}]_{ua,jb}&=0\\[2pt][M_{0}]_{ua,jb}&=0\\[2pt][M_{-}]_{ua,jb}&=0\end{aligned}$\\[12pt]
CO--CO & $\displaystyle\begin{aligned}[t][M_{+}]_{ui,vj}&=0\\[2pt][M_{0}]_{ui,vj}&=\frac{K_{ui,vj}^{\beta\beta,\beta\beta} + \delta_{ij} f_{uv}^{\beta} - \delta_{uv} f_{ji}^{\beta}}{2}\\[2pt][M_{-}]_{ui,vj}&=K_{ui,vj}^{\mathrm{TC}} + \delta_{ij} f_{uv}^{\beta} - \delta_{uv} f_{ji}^{\alpha}\end{aligned}$\\[12pt]
OC--OC & $\displaystyle\begin{aligned}[t][M_{+}]_{iu,jv}&=0\\[2pt][M_{0}]_{iu,jv}&=0\\[2pt][M_{-}]_{iu,jv}&=0\end{aligned}$\\[12pt]
CO--OV & $\displaystyle\begin{aligned}[t][M_{+}]_{ui,bv}&=0\\[2pt][M_{0}]_{ui,bv}&=- \frac{K_{ui,bv}^{\beta\beta,\alpha\alpha}}{2}\\[2pt][M_{-}]_{ui,bv}&=K_{ui,bv}^{\mathrm{TC}}\end{aligned}$\\[12pt]
OC--VO & $\displaystyle\begin{aligned}[t][M_{+}]_{iu,vb}&=0\\[2pt][M_{0}]_{iu,vb}&=0\\[2pt][M_{-}]_{iu,vb}&=0\end{aligned}$\\[12pt]
OV--CO & $\displaystyle\begin{aligned}[t][M_{+}]_{au,vj}&=0\\[2pt][M_{0}]_{au,vj}&=- \frac{K_{au,vj}^{\alpha\alpha,\beta\beta}}{2}\\[2pt][M_{-}]_{au,vj}&=K_{au,vj}^{\mathrm{TC}}\end{aligned}$\\[12pt]
VO--OC & $\displaystyle\begin{aligned}[t][M_{+}]_{ua,jv}&=0\\[2pt][M_{0}]_{ua,jv}&=0\\[2pt][M_{-}]_{ua,jv}&=0\end{aligned}$\\[12pt]
OV--OV & $\displaystyle\begin{aligned}[t][M_{+}]_{au,bv}&=0\\[2pt][M_{0}]_{au,bv}&=\frac{K_{au,bv}^{\alpha\alpha,\alpha\alpha} - \delta_{ab} f_{vu}^{\alpha} + \delta_{uv} f_{ab}^{\alpha}}{2}\\[2pt][M_{-}]_{au,bv}&=K_{au,bv}^{\mathrm{TC}} - \delta_{ab} f_{vu}^{\alpha} + \delta_{uv} f_{ab}^{\beta}\end{aligned}$\\[12pt]
VO--VO & $\displaystyle\begin{aligned}[t][M_{+}]_{ua,vb}&=0\\[2pt][M_{0}]_{ua,vb}&=0\\[2pt][M_{-}]_{ua,vb}&=0\end{aligned}$\\[12pt]
\end{longtable}
\endgroup

\begingroup
\setstretch{1}
\fontsize{9.5}{12}\selectfont
\setlength{\tabcolsep}{6pt}
\setlength{\LTcapwidth}{\textwidth}
\begin{longtable}{@{}p{.15\textwidth}p{.82\textwidth}@{}}
\caption{Matrix Elements of $M_+$, $M_0$, and $M_-$ between Excitation and Deexcitation Blocks without OO Transitions, Based on EOM1}\label{tab:M_lower_B1}\\
\toprule
Blocks & Matrix elements\\
\midrule
\endfirsthead
\toprule
Blocks & Matrix elements\\
\midrule
\endhead
\midrule
\endfoot
\bottomrule
\endlastfoot
CV--VC & $\displaystyle\begin{aligned}[t][M_{+}]_{ai,jb}&=- K_{ai,jb}^{\mathrm{TC}}\\[2pt][M_{0}]_{ai,jb}&=- K_{ai,jb}^{\mathrm{TC}}\\[2pt][M_{-}]_{ai,jb}&=- K_{ai,jb}^{\mathrm{TC}}\end{aligned}$\\[12pt]
VC--CV & $\displaystyle\begin{aligned}[t][M_{+}]_{ia,bj}&=0\\[2pt][M_{0}]_{ia,bj}&=0\\[2pt][M_{-}]_{ia,bj}&=0\end{aligned}$\\[12pt]
CV--OC & $\displaystyle\begin{aligned}[t][M_{+}]_{ai,jv}&=- K_{ai,jv}^{\mathrm{TC}}\\[2pt][M_{0}]_{ai,jv}&=\frac{K_{ai,jv}^{\alpha\alpha,\beta\beta} - K_{ai,jv}^{\beta\beta,\beta\beta}}{2}\\[2pt][M_{-}]_{ai,jv}&=0\end{aligned}$\\[12pt]
VC--CO & $\displaystyle\begin{aligned}[t][M_{+}]_{ia,vj}&=0\\[2pt][M_{0}]_{ia,vj}&=0\\[2pt][M_{-}]_{ia,vj}&=0\end{aligned}$\\[12pt]
OC--CV & $\displaystyle\begin{aligned}[t][M_{+}]_{iu,bj}&=0\\[2pt][M_{0}]_{iu,bj}&=0\\[2pt][M_{-}]_{iu,bj}&=0\end{aligned}$\\[12pt]
CO--VC & $\displaystyle\begin{aligned}[t][M_{+}]_{ui,jb}&=0\\[2pt][M_{0}]_{ui,jb}&=\frac{K_{ui,jb}^{\beta\beta,\alpha\alpha} - K_{ui,jb}^{\beta\beta,\beta\beta}}{2}\\[2pt][M_{-}]_{ui,jb}&=- K_{ui,jb}^{\mathrm{TC}}\end{aligned}$\\[12pt]
CO--OC & $\displaystyle\begin{aligned}[t][M_{+}]_{ui,jv}&=0\\[2pt][M_{0}]_{ui,jv}&=- \frac{K_{ui,jv}^{\beta\beta,\beta\beta}}{2}\\[2pt][M_{-}]_{ui,jv}&=0\end{aligned}$\\[12pt]
OC--CO & $\displaystyle\begin{aligned}[t][M_{+}]_{iu,vj}&=0\\[2pt][M_{0}]_{iu,vj}&=0\\[2pt][M_{-}]_{iu,vj}&=0\end{aligned}$\\[12pt]
CV--VO & $\displaystyle\begin{aligned}[t][M_{+}]_{ai,vb}&=- K_{ai,vb}^{\mathrm{TC}}\\[2pt][M_{0}]_{ai,vb}&=\frac{- K_{ai,vb}^{\alpha\alpha,\alpha\alpha} + K_{ai,vb}^{\beta\beta,\alpha\alpha}}{2}\\[2pt][M_{-}]_{ai,vb}&=0\end{aligned}$\\[12pt]
VC--OV & $\displaystyle\begin{aligned}[t][M_{+}]_{ia,bv}&=0\\[2pt][M_{0}]_{ia,bv}&=0\\[2pt][M_{-}]_{ia,bv}&=0\end{aligned}$\\[12pt]
VO--CV & $\displaystyle\begin{aligned}[t][M_{+}]_{ua,bj}&=0\\[2pt][M_{0}]_{ua,bj}&=0\\[2pt][M_{-}]_{ua,bj}&=0\end{aligned}$\\[12pt]
OV--VC & $\displaystyle\begin{aligned}[t][M_{+}]_{au,jb}&=0\\[2pt][M_{0}]_{au,jb}&=\frac{- K_{au,jb}^{\alpha\alpha,\alpha\alpha} + K_{au,jb}^{\alpha\alpha,\beta\beta}}{2}\\[2pt][M_{-}]_{au,jb}&=- K_{au,jb}^{\mathrm{TC}}\end{aligned}$\\[12pt]
CO--VO & $\displaystyle\begin{aligned}[t][M_{+}]_{ui,vb}&=0\\[2pt][M_{0}]_{ui,vb}&=\frac{K_{ui,vb}^{\beta\beta,\alpha\alpha} - \delta_{uv} f_{bi}^{\beta}}{2}\\[2pt][M_{-}]_{ui,vb}&=- \delta_{uv} f_{bi}^{\alpha}\end{aligned}$\\[12pt]
OC--OV & $\displaystyle\begin{aligned}[t][M_{+}]_{iu,bv}&=0\\[2pt][M_{0}]_{iu,bv}&=0\\[2pt][M_{-}]_{iu,bv}&=0\end{aligned}$\\[12pt]
VO--CO & $\displaystyle\begin{aligned}[t][M_{+}]_{ua,vj}&=0\\[2pt][M_{0}]_{ua,vj}&=0\\[2pt][M_{-}]_{ua,vj}&=0\end{aligned}$\\[12pt]
OV--OC & $\displaystyle\begin{aligned}[t][M_{+}]_{au,jv}&=0\\[2pt][M_{0}]_{au,jv}&=\frac{K_{au,jv}^{\alpha\alpha,\beta\beta} + \delta_{uv} f_{aj}^{\alpha}}{2}\\[2pt][M_{-}]_{au,jv}&=\delta_{uv} f_{aj}^{\beta}\end{aligned}$\\[12pt]
OV--VO & $\displaystyle\begin{aligned}[t][M_{+}]_{au,vb}&=0\\[2pt][M_{0}]_{au,vb}&=- \frac{K_{au,vb}^{\alpha\alpha,\alpha\alpha}}{2}\\[2pt][M_{-}]_{au,vb}&=0\end{aligned}$\\[12pt]
VO--OV & $\displaystyle\begin{aligned}[t][M_{+}]_{ua,bv}&=0\\[2pt][M_{0}]_{ua,bv}&=0\\[2pt][M_{-}]_{ua,bv}&=0\end{aligned}$\\[12pt]
\end{longtable}
\endgroup

\begingroup
\setstretch{1}
\fontsize{9}{11}\selectfont
\setlength{\tabcolsep}{6pt}
\setlength{\LTcapwidth}{\textwidth}
\begin{longtable}{@{}p{.15\textwidth}p{.82\textwidth}@{}}
\caption{Matrix Elements of $M_+$, $M_0$, and $M_-$ Involving OO Transitions, Based on EOM1}\label{tab:M_lower_OO}\\
\toprule
Blocks & Matrix elements\\
\midrule
\endfirsthead
\toprule
Blocks & Matrix elements\\
\midrule
\endhead
\midrule
\endfoot
\bottomrule
\endlastfoot
CV--OO & $\displaystyle\begin{aligned}[t][M_{+}]_{ai,vw}&=- K_{ai,vw}^{\mathrm{TC}}\\[1pt][M_{0}]_{ai,vw}&=0\\[1pt][M_{-}]_{ai,vw}&=K_{ai,vw}^{\mathrm{TC}}\end{aligned}$\\[0pt]
VC--OO & $\displaystyle\begin{aligned}[t][M_{+}]_{ia,vw}&=0\\[1pt][M_{0}]_{ia,vw}&=0\\[1pt][M_{-}]_{ia,vw}&=0\end{aligned}$\\[0pt]
OO--CV & $\displaystyle\begin{aligned}[t][M_{+}]_{tu,bj}&=0\\[1pt][M_{0}]_{tu,bj}&=\frac{\delta_{tu} \left(f_{jb}^{\alpha} - f_{jb}^{\beta}\right)}{2}\\[1pt][M_{-}]_{tu,bj}&=K_{tu,bj}^{\mathrm{TC}}\end{aligned}$\\[0pt]
OO--VC & $\displaystyle\begin{aligned}[t][M_{+}]_{tu,jb}&=0\\[1pt][M_{0}]_{tu,jb}&=\frac{\delta_{tu} \left(- f_{bj}^{\alpha} + f_{bj}^{\beta}\right)}{2}\\[1pt][M_{-}]_{tu,jb}&=- K_{tu,jb}^{\mathrm{TC}}\end{aligned}$\\[0pt]
CO--OO & $\displaystyle\begin{aligned}[t][M_{+}]_{ui,vw}&=0\\[1pt][M_{0}]_{ui,vw}&=- \frac{\delta_{uv} f_{wi}^{\beta}}{2}\\[1pt][M_{-}]_{ui,vw}&=K_{ui,vw}^{\mathrm{TC}} - \delta_{uv} f_{wi}^{\alpha}\end{aligned}$\\[0pt]
OC--OO & $\displaystyle\begin{aligned}[t][M_{+}]_{iu,vw}&=0\\[1pt][M_{0}]_{iu,vw}&=0\\[1pt][M_{-}]_{iu,vw}&=0\end{aligned}$\\[0pt]
OO--CO & $\displaystyle\begin{aligned}[t][M_{+}]_{tu,vj}&=0\\[1pt][M_{0}]_{tu,vj}&=- \frac{\delta_{tu} f_{jv}^{\beta}}{2}\\[1pt][M_{-}]_{tu,vj}&=K_{tu,vj}^{\mathrm{TC}} - \delta_{tv} f_{ju}^{\alpha}\end{aligned}$\\[0pt]
OO--OC & $\displaystyle\begin{aligned}[t][M_{+}]_{tu,jv}&=0\\[1pt][M_{0}]_{tu,jv}&=\frac{\delta_{tu} f_{vj}^{\beta}}{2}\\[1pt][M_{-}]_{tu,jv}&=\delta_{uv} f_{tj}^{\beta}\end{aligned}$\\[0pt]
OV--OO & $\displaystyle\begin{aligned}[t][M_{+}]_{au,vw}&=0\\[1pt][M_{0}]_{au,vw}&=\frac{\delta_{uw} f_{av}^{\alpha}}{2}\\[1pt][M_{-}]_{au,vw}&=K_{au,vw}^{\mathrm{TC}} + \delta_{uw} f_{av}^{\beta}\end{aligned}$\\[0pt]
VO--OO & $\displaystyle\begin{aligned}[t][M_{+}]_{ua,vw}&=0\\[1pt][M_{0}]_{ua,vw}&=0\\[1pt][M_{-}]_{ua,vw}&=0\end{aligned}$\\[0pt]
OO--OV & $\displaystyle\begin{aligned}[t][M_{+}]_{tu,bv}&=0\\[1pt][M_{0}]_{tu,bv}&=\frac{\delta_{tu} f_{vb}^{\alpha}}{2}\\[1pt][M_{-}]_{tu,bv}&=K_{tu,bv}^{\mathrm{TC}} + \delta_{uv} f_{tb}^{\beta}\end{aligned}$\\[0pt]
OO--VO & $\displaystyle\begin{aligned}[t][M_{+}]_{tu,vb}&=0\\[1pt][M_{0}]_{tu,vb}&=- \frac{\delta_{tu} f_{bv}^{\alpha}}{2}\\[1pt][M_{-}]_{tu,vb}&=- \delta_{tv} f_{bu}^{\alpha}\end{aligned}$\\[0pt]
OO--OO & $\displaystyle\begin{aligned}[t][M_{+}]_{tu,vw}&=0\\[1pt][M_{0}]_{tu,vw}&=0\\[1pt][M_{-}]_{tu,vw}&=K_{tu,vw}^{\mathrm{TC}} - \delta_{tv} f_{wu}^{\alpha} + \delta_{uw} f_{tv}^{\beta}\end{aligned}$\\[0pt]
\end{longtable}
\endgroup
OO denotes the full ordered $(t,u)$ product space in this table, including TT.

\bibliographystyle{unsrtnat}
\bibliography{references}
\end{document}